\documentclass[12pt,preprint]{aastex}

\shorttitle{Abundances in M10}
\shortauthors{Haynes et al.}

\newcommand\iso[2]{$^{\rm #1}$#2}

\begin{document}

\title{Chemical Analysis of Five Red Giants in the Globular Cluster M10 (NGC 
6254)}

\author{
Sharina Haynes\altaffilmark{1},
Geoffrey Burks\altaffilmark{1},
Christian I. Johnson\altaffilmark{2}, and
Catherine A. Pilachowski\altaffilmark{2}
}

\altaffiltext{1}{Center of Excellence in Information Systems, Tennessee
State University, 3500 John Merritt Blvd, Box 9501, Holland Hall Room 311, 
Nashville, Tennessee 37209, USA; burks@coe.tsuniv.edu}

\altaffiltext{2}{Department of Astronomy, Indiana University,
Swain West 319, 727 East Third Street, Bloomington, IN 47405--7105, USA;
cijohnson@astro.indiana.edu; catyp@astro.indiana.edu}

\begin{abstract}

We have determined Al, $\alpha$, Fe--peak, and neutron capture elemental
abundances for five red giant branch (RGB) stars in the Galactic globular
cluster M10.  Abundances were determined using equivalent width analyses 
of moderate resolution (R$\sim$15,000) spectra obtained with the Hydra 
multifiber positioner and bench spectrograph on the WIYN telescope.  The data
sample the upper RGB from the luminosity level near the horizontal branch
to about 0.5 mag below the RGB tip.  We find in agreement with previous studies
that M10 is moderately metal--poor with [Fe/H]=--1.45 ($\sigma$=0.04).  All
stars appear enhanced in Al with $\langle$[Al/Fe]$\rangle$=+0.33 
($\sigma$=0.19), but no stars have [Al/Fe]$\ga$+0.55.  We find the $\alpha$
elements to be enhanced by +0.20 to +0.40 dex and the Fe--peak elements to have
[el/Fe]$\sim$0, which are consistent with predictions from type II SNe ejecta.
Additionally, the cluster appears to be r--process rich with 
$\langle$[Eu/La]$\rangle$=+0.41.

\end{abstract}


\section{INTRODUCTION}

Although few chemical analysis studies of M10 exist, the general consensus is 
that this cluster exhibits all of the classical characteristics observed 
in other Galactic globular clusters.  With a metallicity of 
[Fe/H]$\approx$--1.5 (Kraft et al. 1995), M10 lies near the median metallicity 
distribution for halo globular clusters (Laird et al. 1988).  Small sample 
(N$\la$15) analyses of red giant branch (RGB) stars in this cluster have 
revealed it to have [$\alpha$/Fe]$\sim$+0.30 and [el/Fe]$\sim$0 for Fe--peak 
elements (Kraft et al. 1995; Mishenina et al. 2003).  These values are 
consistent with the current generation of M10 stars having been polluted by 
the ejecta of type II supernovae (SNe) without significant contributions from 
type Ia SNe.

While it has long been known that nearly all globular cluster giants show
star--to--star variations of the light elements (A$\la$27), the source of many 
of these anomalies has yet to be determined.  Numerous observations of
globular cluster stars from the main sequence to above the RGB luminosity bump
have revealed declining [C/Fe] and increasing [N/Fe] ratios as a function of 
increasing luminosity (e.g., see reviews by Kraft et al. 1994; Gratton et al. 
2004; Carretta 2008).  These observations show clear evidence of CN--cycle 
products being brought to the surface and are a confirmation of first 
dredge--up predictions (Iben 1964).  Smith \& Fulbright (1997) and Smith et al.
(2005) have verified this trend in M10 as well as a CN band anticorrelation
with [O/Fe] for stars at various RGB luminosities.  However, the large spread
in [N/Fe] of about 1.0 dex found by Smith et al. (2005) in M10 stars may be 
evidence for primordial variations superimposed on in situ mixing.

The C and N abundance anomalies are known to exist in both globular cluster
and field giants, but that likeness does not extend to the well documented
O/Na, Mg/Al, and O/F anticorrelations and Na/Al correlation seen solely in 
globular cluster stars (e.g., Gratton et al. 2004).  These abundance 
relationships are clear signs of proton--capture nucleosynthesis, but where 
these processes are operating is still a mystery.  Kraft et al. (1995) examined
the O/Na anticorrelations of 15 bright giants in M10 along with M3 and M13, 
which are all globular clusters of similar metallicity ([Fe/H]$\approx$--1.5), 
because M10 and M13 have extremely blue horizontal branches (HB) but M3 has a 
uniform distribution of blue HB, RR Lyrae, and red HB stars.  The study showed 
that M10 appears to be an intermediate case in terms of O depletion and Na 
enhancement in that the average [O/Fe] is lower in M10 than in M3, but no M10 
giants were super O--poor (i.e., [O/Fe]$<$--0.6), suggesting the process 
driving O depletion does not itself determine HB morphology.

In this paper we have examined five additional RGB stars in M10 that are 
located above the luminosity of the horizontal branch but below the RGB tip.
We have derived Al, $\alpha$, Fe--peak, and heavy element abundances to 
examine how M10 fits into context with other globular clusters of similar
metallicity and HB morphology.  

\section{OBSERVATIONS AND REDUCTIONS}

The observation of cluster giants were obtained using the Hydra multifiber
positioner and bench spectrograph on the 3.5 meter WIYN telescope at Kitt 
Peak National Observatory in May, 2000.  The observations consisted of three,
3,000 second exposures with the 200 $\mu$m red fiber bundle.  The 316 line 
mm$^{\rm -1}$ echelle grating and red camera provided a resolution of R
($\lambda$/$\Delta$$\lambda$)$\sim$15,000 at 6650~\AA, with wavelength 
coverage extending from approximately 6460--6860~\AA.  

Target stars and photometry were taken from the photometric survey by Arp 
(1955) and astrometry was taken from the USNO Image and Catalogue 
Archive.\footnote{http://www.nofs.navy.mil/data/fchpix/}
Sample selection focused on observing stars with V magnitudes brighter than 
the HB and extending up to the RGB tip.  However, the final sample only 
includes stars with magnitudes up to about 0.5 mag below the RGB tip.  The 
Hydra configuration allowed for fiber placement on 22 objects, but only 7 of 
those 22 had sufficient signal--to--noise (S/N) for reliable abundance 
determinations.  Two of the remaining program stars (IV--44 \& IV--87) were 
found to have wavelength shifts and H$\alpha$ profiles inconsistent with being 
both cluster members and low surface gravity RGB stars.  Comparison with the 
proper motion study by Chen et al. (2000) reveals that both of these stars 
have proper motions inconsistent with other cluster members.

\begin{figure}
\epsscale{.8}
\plotone{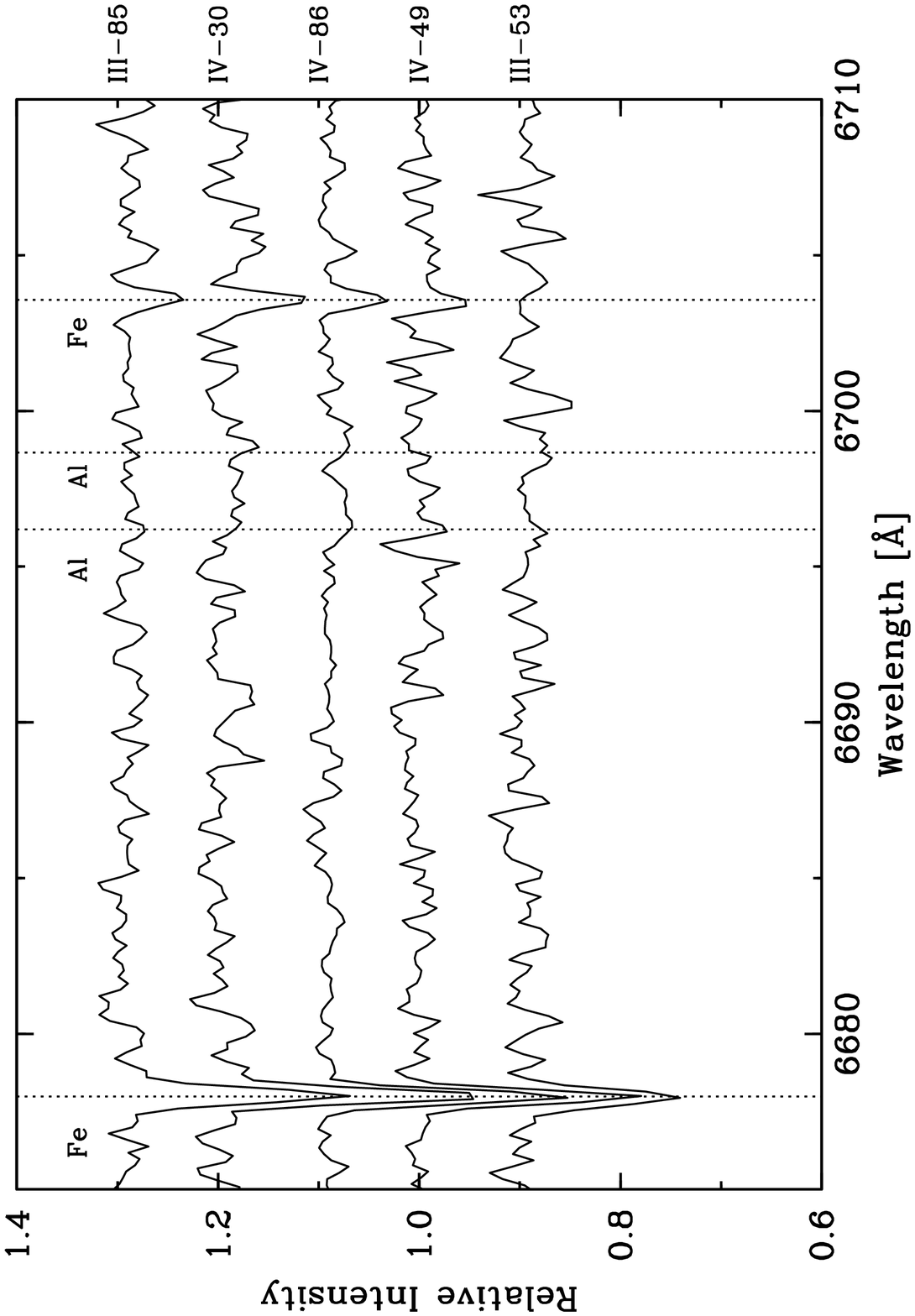}
\caption{Spectra from all five program stars are shown above with a pair of 
Fe I and Al I lines indicated for reference.  The spectra have been offset
for display purposes.}
\label{f1}
\end{figure}

The IRAF\footnote{IRAF is distributed by the National Optical Astronomy 
Observatory, which is operated by AURA, Inc., under cooperative agreement with 
the National Science Foundation.} task {\it ccdproc} was used to trim the bias 
overscan region and apply the bias level correction.  The IRAF routine {\it 
dohydra} was employed to apply the flat field correction, linearize the 
wavelength scale, correct for scattered light, remove cosmic rays, subtract 
the sky, and extract the one--dimensional spectra.  Typical S/N ratios of
individual spectra are 25--50 with co--added spectra having S/N ratios of about
50--75.  A sample spectral region for all five program stars is shown in 
Figure \ref{f1}.

\section{ANALYSIS}

\subsection{Model Stellar Atmospheres}

The significant differential reddening associated with M10 (e.g., von Braun et 
al. 2002) makes effective temperature (T$_{\rm eff}$) and surface gravity 
(log g) estimates based on color and photometric indices difficult.  
Therefore, we employed an iterative method to obtain T$_{\rm eff}$ by 
removing Fe I abundance trends as a function of excitation potential and 
microturbulence (V$_{\rm t}$) by removing Fe I abundance trends as a function 
of reduced width [log(EW/$\lambda$)].  Surface gravity was obtained by 
enforcing ionization equilibrium between Fe I and Fe II because ionized Fe 
lines in these cool giants are more sensitive to changes in log g than neutral 
lines (e.g., Johnson \& Pilachowksi 2006; their Table 3).  Despite the fact
that only one Fe II line (6516~\AA) was available for analysis, we used the 
ionization equilibrium method because of the potentially large and variable 
uncertainties in determining bolometric absolute magnitudes.

All initial models assumed a metallicity of [Fe/H] = --1.50, which is 
consistent with previous spectroscopic [Fe/H] estimates (e.g., Kraft et al. 
1995; Mishenina et al. 2003).  The model stellar atmospheres (without 
convective overshoot) were created by interpolating in the ATLAS 
grid\footnote{Kurucz model atmosphere grids can be downloaded from 
http://cfaku5.cfa.harvard.edu/grids.html.} (Castelli et al. 1997).  The 
temperature range of our observations covers 4450 $\le$ T$_{\rm eff}$ $\le$ 
4750 corresponding to surface gravity values of about 1.20 $\le$ log g $\le$ 
1.85.  Our adopted values of T$_{\rm eff}$ and log g are in reasonable 
agreement with position on the color--magnitude diagram as estimated from 
V and B--V photometry.  A summary of our adopted model atmosphere parameters 
and associated photometry is provided in Table 1.

\subsection{Equivalent Width Analyses}

All abundances were determined by measuring equivalent widths using the {\it
splot} package in IRAF.  Given the moderate resolution and S/N of our spectra,
we restricted measurements to isolated lines that did not suffer significant
blending problems and which had equivalent widths $\ga$10 m\AA.  Suitable
lines were chosen via comparison with a high S/N, high resolution Arcturus
spectrum\footnote{The Arcturus Atlas can be downloaded from the NOAO Digital 
Library at http://www.noao.edu/dpp/library.html.}, which also served as a 
reference aiding continuum placement.  The final linelist including all
measured equivalent widths is given in Table 2, with atomic parameters taken 
from Johnson \& Pilachowski (2006).  

While most abundances were calculated using the {\it abfind} driver in the 2002 
version of the LTE line analysis code MOOG (Sneden 1973), the elements Sc and
Eu required a modified approach.  These spectral features may be sensitive
to hyper--fine splitting resulting from spin--orbit coupling and Eu has the 
added complication of having two naturally occurring stable isotopes 
(\iso{151}{Eu} and \iso{153}{Eu}).  Both of these effects can cause line 
broadening that will force single--line equivalent width measurements to 
overestimate the abundances.  Therefore, we used the {\it blends} driver in 
MOOG with linelists including hyperfine and/or isotopic data from Prochaska \& 
McWilliam (2000) for Sc and C. Sneden (private communication, 2006) for Eu.  
While the 6774~\AA\ La II line may also be sensitive to hyper--fine splitting, 
no known linelist exists in the literature for this transition.  However, the 
typically small equivalent widths of this line ($\la$30 m\AA) suggest 
additional broadening will not affect La abundances too severely.

\section{RESULTS AND DISCUSSION}

\subsection{Al Abundances}

We have determined at least upper limits of [Al/Fe] for five giants with the 
cluster having $\langle$[Al/Fe]$\rangle$=+0.33 ($\sigma$=0.19) and a full 
range of 0.50 dex.  Both the star--to--star dispersion and average 
[Al/Fe] ratios are in agreement with observations of other Galactic globular 
clusters of similar metallicity (e.g., Kraft et al. 1998; Sneden et al. 2004; 
Cohen \& Mel{\'e}ndez 2005; Johnson et al. 2005; Yong et al. 2005); however, 
the highest [Al/Fe] ratio found in our sample is about a factor of three 
smaller than the $>$+1.0 dex ratios observed in M3 and M13 (Pilachowski et al. 
1996; Sneden et al. 2004; Johnson et al. 2005; Cohen \& Mel{\'e}ndez 2005), 
which possess similar metallicity and, in the case of M13, a similar HB 
morphology.  This may be due to our small sample size coupled with 
observations of stars well below the RGB tip, where additional Al enhancement 
due to extra in situ mixing may be operating (e.g., Denissenkov \& VandenBerg 
2003).  Kraft et al. (1995) found M10 to be an intermediate case between M3 
and M13 with regard to the amount of O depletion and Na enhancement and 
therefore given the likely Na--Al correlation present in this cluster one 
would not expect [Al/Fe] values much greater than about +0.80 dex.  A complete 
list of our determined abundances for Al and all other elements is provided in 
Table 3.

It has been shown that [Fe/H] determinations based on Fe I lines in metal--poor
stars suffer from larger LTE departure effects than their metal--rich 
counterparts because of overionization due to decreased UV line blocking 
(e.g., see review by Asplund 2005).  Correcting for this effect would drive 
the [Fe/H] abundance up, perhaps by as much as $\sim$+0.30 dex at [Fe/H]=--3 
(Th{\'e}venin \& Idiart 1999, but see also Gratton et al. 1999; Kraft \& Ivans 
2003), and thus decrease the derived [Al/Fe] ratio found here.  While a few 
NLTE studies for Al exist (e.g., Gehren et al. 2004; Andrievsky et al. 2008)
finding offsets of order a few tenths of a dex, the actual Al NLTE correction 
for stars in the metallicity and luminosity regime studied here are mostly 
unknown.  Fortunately, our sample does not vary widely in either metallicity or 
luminosity and any NLTE corrections are likely to be very similar, suggesting 
at least the relative star--to--star dispersion is a real effect.

\begin{figure}
\epsscale{1.00}
\plotone{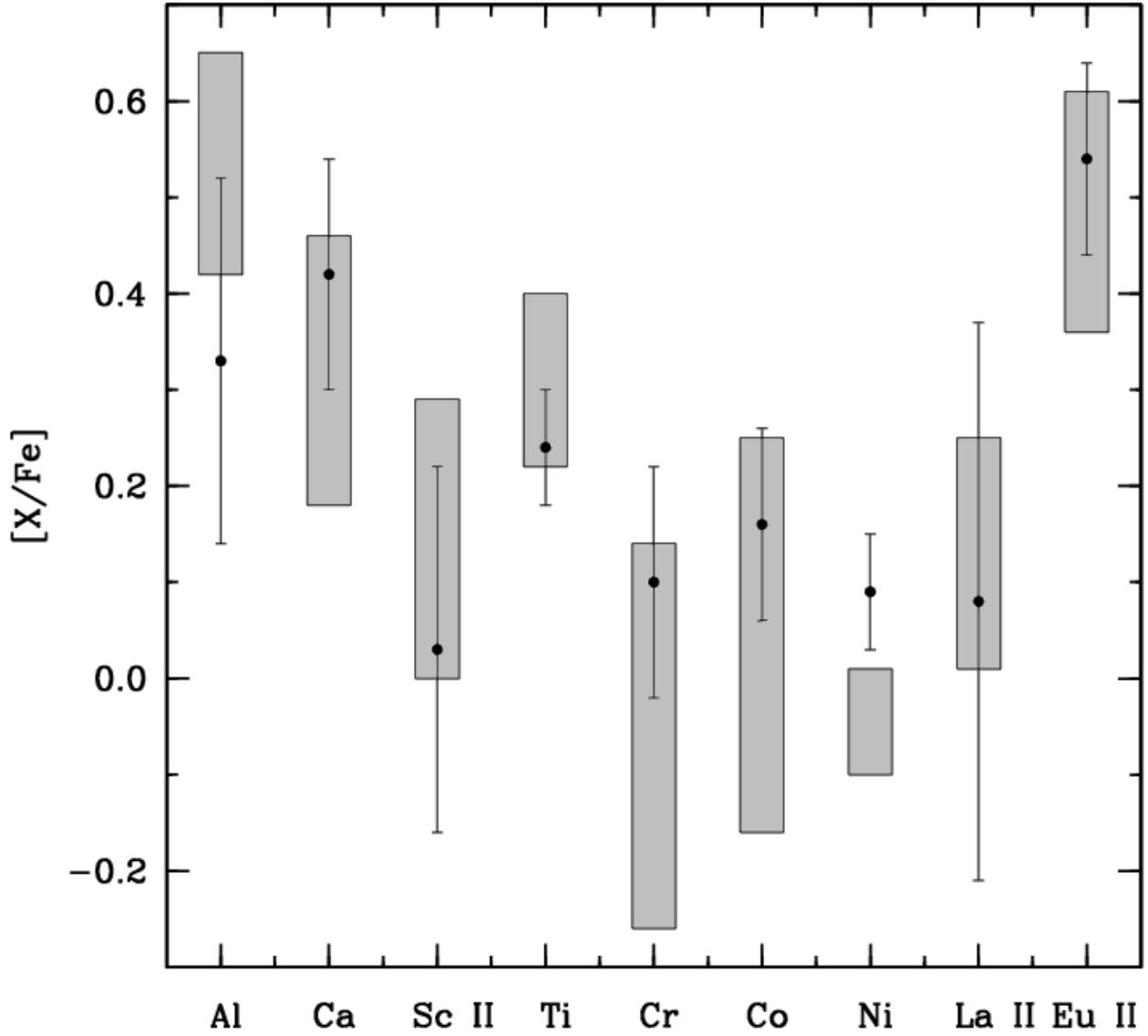}
\caption{Average abundances for various elements in M10 are shown as filled
circles with error bars representing the 1$\sigma$ values.  Similar data for
M12 from Johnson \& Pilachowski (2006) are shown as shaded boxes overlapping
the M10 results.}
\label{f2}
\end{figure}

\begin{figure}
\epsscale{.8}
\plotone{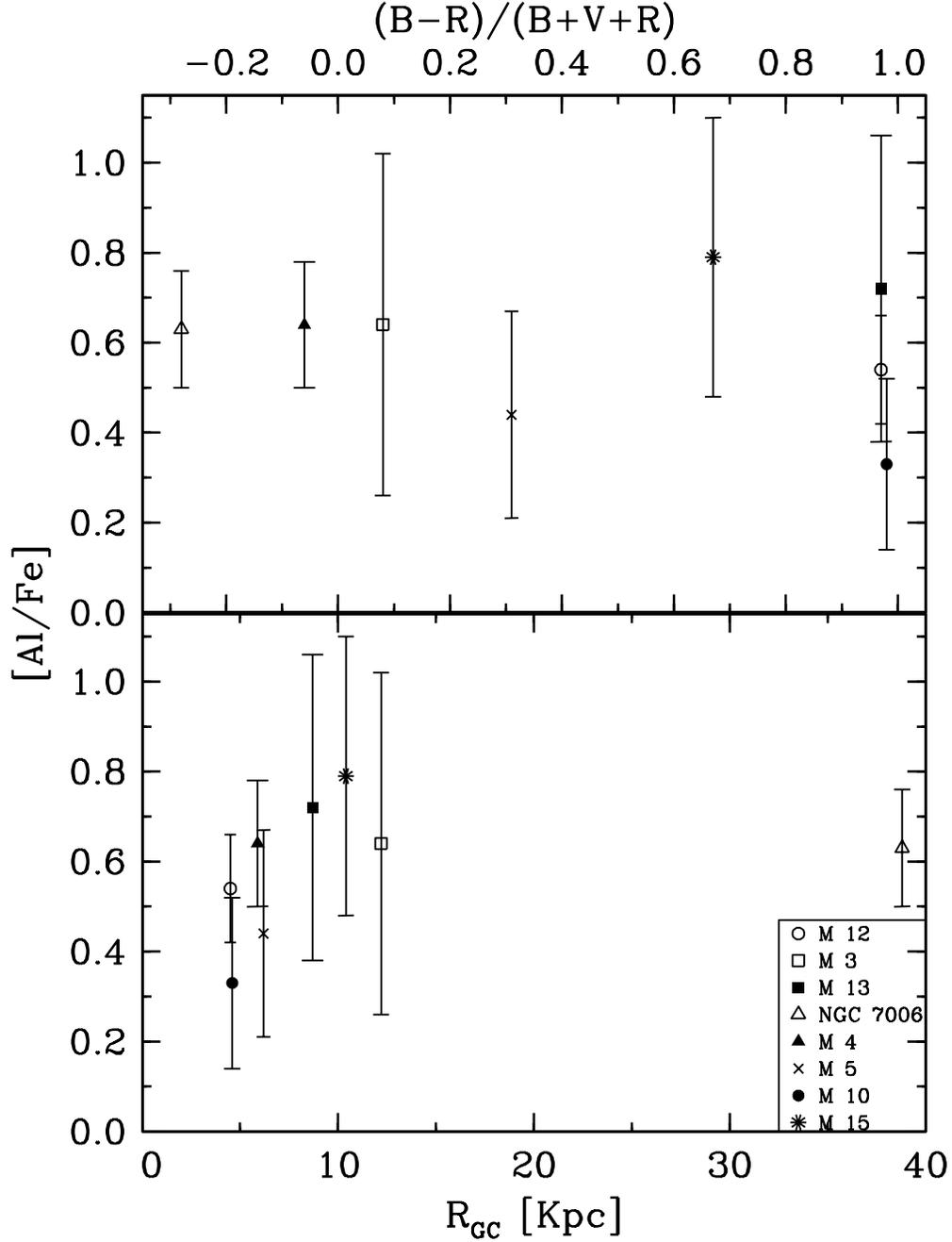}
\caption{The top panel shows [Al/Fe] versus the horizontal branch parameter, 
where a value of +1.0 means only BHB stars, 0.0 is a mix of RHB, BHB, and RR
Lyrae stars, and --1.0 means only RHB stars.  The various symbols indicate the
average [Al/Fe] abundance in a cluster with the error bars showing the 
1$\sigma$ values.  The bottom panel shows [Al/Fe] versus the Galactocentric
distance for each cluster.  The cluster data are from: M12 (Johnson \& 
Pilachowski 2006), M3 and M13 (Johnson et al. 2005), NGC 7006 (Kraft et al.
1998), M4 (Ivans et al. 1999), M5 (Ivans et al. 2001), and M15 (Sneden et al.
1997).}
\label{f3}
\end{figure}

In Figure \ref{f2} we compare abundances of various elements in M10 versus
those in the similar cluster M12.  The [Al/Fe] abundances for both clusters
are comparable and each displays a modest star--to--star dispersion.  Given 
that the scatter is about a factor of two larger than those observed in 
the Fe--peak and $\alpha$ elements, it is likely that the Al distribution is
real and not an artifact of observational uncertainty.  To see how M10 fits
into the context of other Galactic globular clusters, we have plotted [Al/Fe]
as a function of both horizontal branch ratio (HBR) and galactocentric 
distance (R$_{\rm GC}$) for M10 and seven other clusters in Figure \ref{f3}.
The top panel suggests there is no significant relation between HBR and 
either the average [Al/Fe] ratio or the star--to--star dispersion.  However,
it should be noted that Carretta et al. (2007) do find a relationship between 
the extent of O/Mg depletions and Na/Al enhancements and the maximum
temperature of stars located on the zero--age HB.  The bottom panel may 
indicate a trend of increasing cluster average [Al/Fe] with increasing 
galactocentric distance; however, the sample size for each cluster varies 
between less than 10 to nearly 100 stars.  Consequently, M10 does not appear 
to exhibit anomalous [Al/Fe] ratios compared to other globular clusters.

\subsection{$\alpha$, Fe--Peak, and Heavy Elements}

Nearly all globular clusters with [Fe/H]$<$--1 have [$\alpha$/Fe]$\sim$+0.30 to
+0.50, solar Fe--peak to Fe ratios, and are r--process rich (e.g., Gratton et 
al. 2004).  The star--to--star scatter present is usually $\la$0.10 dex for 
the $\alpha$ and Fe--peak elements and $\sim$0.30--0.50 dex for the neutron 
capture elements, which is still significantly less than the 0.50--1.00 dex 
variations seen in light elements such as O, Na, and Al.  In M10 we find
the expected enhancement and small star--to--star dispersion of the two 
$\alpha$ elements Ca and Ti with $\langle$[Ca/Fe]$\rangle$=+0.42 
($\sigma$=0.12) and $\langle$[Ti/Fe]$\rangle$=+0.24 ($\sigma$=0.06).  These
values are consistent with the results from Kraft et al. (1995) that
found $\langle$[Ca/Fe]$\rangle$=+0.29 ($\sigma$=0.07) and 
$\langle$[Ti/Fe]$\rangle$=+0.21 ($\sigma$=0.12) for a set of 10 other upper
RGB stars in this cluster.  The proxy Fe--peak elements Sc and Ni exhibit
near solar abundance ratios in all stars with cluster average values of 
$\langle$[Sc/Fe]$\rangle$=+0.03 ($\sigma$=0.19) and 
$\langle$[Ni/Fe]$\rangle$=+0.09 ($\sigma$=0.06), which are roughly consistent
with Kraft et al. (1995).  These abundances patterns are mirrored in M12
(see Figure \ref{f2}), but with M10 showing a smaller range of [Cr/Fe] and 
[Co/Fe] abundances.  The combination of $\alpha$ enhancement and near solar
Fe--peak ratios is consistent with this cluster being primarily polluted by
the ejecta of type II SNe (e.g., Woosley \& Weaver 1995).

For stars near M10's metallicity, La is produced primarily via the s--process 
in $\sim$1--3 M$_{\odot}$ stars and Eu from the r--process in $\sim$8--10
M$_{\odot}$ stars (e.g., Busso et al. 1999; Truran et al. 2002).  Our derived 
La and Eu abundances are consistent with the picture of massive stars 
producing most of the heavy elements in this cluster with 
$\langle$[La/Fe]$\rangle$=+0.08 ($\sigma$=0.29) and 
$\langle$[Eu/Fe]$\rangle$=+0.54 ($\sigma$=0.10).  Comparing the ratio of 
r-- to s--process elements gives [Eu/La]=+0.41 and implies M10 is slightly
more r--process rich than the average globular cluster.  However, this 
value is within the 1$\sigma$ range of $\langle$[Eu/Ba,La]$\rangle$=+0.23
($\sigma$=0.21) found by Gratton et al. (2004) after combining data from the
literature on 28 globular clusters.  A larger sample size of M10 stars is
likely to decrease the star--to--star scatter observed in our La and Eu 
sample but will probably not change the result that the cluster is r--process
rich.

\section{SUMMARY}

We have determined abundances of the light element Al as well as several
$\alpha$, Fe--peak, and heavy elements in five M10 red giants using moderate
resolution spectroscopy (R$\sim$15,000) obtained with the Hydra multifiber 
spectrograph on the WIYN telescope.   The data sample the upper RGB with 
luminosities ranging from above the level of the HB to about 0.5 mag below
the RGB tip.  Model atmosphere parameters were determined via spectroscopic 
procedures relying on abundances from equivalent width analyses.

Our results are in agreement with previous studies that M10 is metal--poor 
with [Fe/H]=--1.45 and has a very small metallicity spread ($\sigma$=0.04).
Al abundances indicate that while cluster stars maintain supersolar [Al/Fe]
values, there is a paucity of high--Al stars (i.e., [Al/Fe]$\ga$+1.0).  This
result corroborates the O and Na data from Kraft et al. (1995) who found no 
stars with [O/Fe]$<$--0.6, despite the cluster's similarity to M13 which has
several super O--poor/high--Al RGB stars.  The modest average Al enhancement 
of [Al/Fe]=+0.33 may be a consequence of its galactocentric distance of 
$\sim$5 Kpc because comparison with several other similar metallicity globular 
clusters at different R$_{\rm GC}$ shows a possible trend of increasing 
$\langle$[Al/Fe]$\rangle$ with increasing R$_{\rm GC}$.

We find all stars to have enhancements in [Ca/Fe] and [Ti/Fe] by about +0.20
to +0.40 dex and [el/Fe]$\sim$0 for Fe--peak elements.  These data suggest the 
current generation of M10 stars were heavily polluted with the by--products of 
type II SNe without significant type Ia contributions, which would result in
lower [$\alpha$/Fe] ratios.  The neutron capture elements also suggest 
enrichment from massive stars because the cluster appears to be quite 
r--process rich with $\langle$[Eu/La]$\rangle$=+0.41.

\acknowledgements

We are grateful to Diane Harmer for obtaining all observations used in this
paper.  We would like to thank the NSF REU program for supporting SH via 
AST--0453437.  Support of the College of Arts and Sciences at Indiana 
University Bloomington for CIJ is gratefully acknowledged.

\newpage

\newpage

\newpage

\clearpage

\tablenum{1}
\tablecolumns{7}
\tablewidth{0pt}

\begin{deluxetable}{ccccccc}
\tablecaption{Photometry and Model Atmosphere Parameters}
\tablehead{
\colhead{Star\tablenotemark{a}} 	& 	
\colhead{V} 	& 	
\colhead{B$-$V} 	&	
\colhead{T$_{\rm eff}$}	&	
\colhead{log g}	& 
\colhead{[Fe/H]}	&
\colhead{v$_{\rm t}$}	\\
\colhead{}	&
\colhead{}      &
\colhead{}      &
\colhead{(K)}      &
\colhead{(cm s$^{\rm -2}$)}      &
\colhead{Spectroscopy}      &
\colhead{(km s$^{\rm -1}$)} 
}

\startdata
II-85	&	12.76	&	1.37	&	4450	&	1.20	&	$-$1.42	&	1.70	\\
IV-30	&	12.77	&	1.44	&	4450	&	1.20	&	$-$1.47	&	1.90	\\
IV-86	&	13.13	&	1.25	&	4550	&	1.50	&	$-$1.42	&	1.55	\\
III-53	&	13.80	&	1.09	&	4750	&	1.85	&	$-$1.44	&	1.20	\\
IV-49	&	14.47	&	1.10	&	4700	&	1.85	&	$-$1.52	&	1.30	\\
\enddata

\tablenotetext{a}{Identifiers are from Arp (1955).}

\end{deluxetable}

\clearpage

\tablenum{2}
\tablecolumns{9}
\tablewidth{0pt}

\begin{deluxetable}{ccccccccc}
\tablecaption{Linelist and Equivalent Widths\tablenotemark{a,b}}
\tablehead{
\colhead{$\lambda$} 	& 	
\colhead{Element}      &
\colhead{E.P.}      &
\colhead{log gf}      &
\colhead{III$-$53}      &
\colhead{III$-$85}      &
\colhead{IV$-$30}      &
\colhead{IV$-$49}      &
\colhead{IV$-$86} \\
\colhead{(\AA)}	&
\colhead{}      &
\colhead{eV}      &
\colhead{}      &
\colhead{}      &
\colhead{}      &
\colhead{}      &
\colhead{}      &
\colhead{}
}
\startdata
6696.03	&	Al I	&	3.14	&	$-$1.57	&	13	&	12	&	15	&	14	&	26	\\
6698.66	&	Al I	&	3.14	&	$-$1.89	&	\nodata	&	\nodata	&	\nodata	&	\nodata	&	16	\\
6471.68	&	Ca I	&	2.52	&	$-$0.69	&	58	&	112	&	101	&	82	&	98	\\
6499.65	&	Ca I	&	2.52	&	$-$0.82	&	65	&	75	&	108	&	70	&	87	\\
6717.68	&	Ca I	&	2.71	&	$-$0.61	&	\nodata	&	\nodata	&	97	&	\nodata	&	\nodata	\\
6604.60	&	Sc II	&	1.36	&	$-$1.48	&	26	&	49	&	53	&	40	&	63	\\
6554.23	&	Ti I	&	1.44	&	$-$1.16	&	\nodata	&	44	&	27	&	22	&	17	\\
6556.07	&	Ti I	&	1.46	&	$-$1.10	&	14	&	30	&	64	&	17	&	38	\\
6743.12	&	Ti I	&	0.90	&	$-$1.65	&	24	&	43	&	48	&	\nodata	&	42	\\
6559.57	&	Ti II	&	2.05	&	$-$2.30	&	\nodata	&	55	&	40	&	35	&	45	\\
6606.97	&	Ti II	&	2.06	&	$-$2.79	&	\nodata	&	22	&	22	&	16	&	14	\\
6630.03	&	Cr I	&	1.03	&	$-$3.49	&	\nodata	&	\nodata	&	14	&	\nodata	&	15	\\
6475.63	&	Fe I	&	2.56	&	$-$3.01	&	44	&	65	&	74	&	36	&	65	\\
6481.87	&	Fe I	&	2.28	&	$-$3.08	&	\nodata	&	87	&	73	&	\nodata	&	77	\\
6498.95	&	Fe I	&	0.96	&	$-$4.69	&	\nodata	&	96	&	90	&	44	&	68	\\
6533.93	&	Fe I	&	4.56	&	$-$1.36	&	14	&	16	&	\nodata	&	\nodata	&	15	\\
6546.24	&	Fe I	&	2.76	&	$-$1.54	&	85	&	126	&	\nodata	&	95	&	107	\\
6574.25	&	Fe I	&	0.99	&	$-$5.02	&	25	&	65	&	69	&	\nodata	&	45	\\
6592.92	&	Fe I	&	2.73	&	$-$1.47	&	90	&	145	&	132	&	103	&	134	\\
6593.88	&	Fe I	&	2.43	&	$-$2.42	&	65	&	106	&	105	&	\nodata	&	102	\\
6597.57	&	Fe I	&	4.79	&	$-$0.95	&	\nodata	&	\nodata	&	16	&	\nodata	&	19	\\
6608.04	&	Fe I	&	2.28	&	$-$3.96	&	\nodata	&	19	&	\nodata	&	\nodata	&	26	\\
6609.12	&	Fe I	&	2.56	&	$-$2.69	&	50	&	72	&	103	&	39	&	60	\\
6625.02	&	Fe I	&	1.01	&	$-$5.37	&	\nodata	&	47	&	37	&	30	&	37	\\
6627.54	&	Fe I	&	4.55	&	$-$1.58	&	\nodata	&	11	&	\nodata	&	\nodata	&	13	\\
6646.96	&	Fe I	&	2.61	&	$-$3.96	&	\nodata	&	15	&	\nodata	&	\nodata	&	\nodata	\\
6648.12	&	Fe I	&	1.01	&	$-$5.92	&	\nodata	&	27	&	\nodata	&	\nodata	&	15	\\
6677.99	&	Fe I	&	2.69	&	$-$1.35	&	119	&	143	&	160	&	108	&	135	\\
6703.57	&	Fe I	&	2.76	&	$-$3.01	&	25	&	38	&	54	&	\nodata	&	36	\\
6710.32	&	Fe I	&	1.48	&	$-$4.83	&	\nodata	&	30	&	44	&	\nodata	&	31	\\
6726.67	&	Fe I	&	4.61	&	$-$1.07	&	\nodata	&	\nodata	&	27	&	14	&	21	\\
6733.15	&	Fe I	&	4.64	&	-1.48	&	\nodata	&	11	&	\nodata	&	\nodata	&	\nodata	\\
6739.52	&	Fe I	&	1.56	&	$-$4.79	&	\nodata	&	\nodata	&	26	&	\nodata	&	27	\\
6750.16	&	Fe I	&	2.42	&	$-$2.62	&	53	&	98	&	91	&	69	&	91	\\
6806.85	&	Fe I	&	2.73	&	$-$3.10	&	20	&	38	&	34	&	\nodata	&	23	\\
6516.08	&	Fe II	&	2.89	&	$-$3.45	&	\nodata	&	54	&	54	&	35	&	45	\\
6632.47	&	Co I	&	2.28	&	$-$1.85	&	9	&	17	&	\nodata	&	\nodata	&	18	\\
6482.80	&	Ni I	&	1.93	&	$-$2.79	&	\nodata	&	74	&	\nodata	&	40	&	74	\\
6532.88	&	Ni I	&	1.93	&	$-$3.47	&	\nodata	&	37	&	\nodata	&	19	&	24	\\
6586.31	&	Ni I	&	1.95	&	$-$2.81	&	31	&	56	&	42	&	\nodata	&	54	\\
6643.63	&	Ni I	&	1.68	&	$-$2.01	&	98	&	135	&	140	&	84	&	118	\\
6767.78	&	Ni I	&	1.83	&	$-$2.17	&	76	&	102	&	125	&	\nodata	&	91	\\
6772.32	&	Ni I	&	3.66	&	$-$0.96	&	\nodata	&	31	&	\nodata	&	\nodata	&	27	\\
6774.33	&	La II	&	0.13	&	$-$1.75	&	\nodata	&	16	&	29	&	\nodata	&	7	\\
6645.12	&	Eu II	&	1.37	&	$+$0.20	&	\nodata	&	24	&	23	&	18	&	22	\\

\enddata
\tablenotetext{a}{Identifiers are from Arp (1955).}
\tablenotetext{b}{Equivalent widths are given in units of m\AA.}
\end{deluxetable}

\clearpage
\pagestyle{empty}

\tablenum{3}
\tablecolumns{13}
\tablewidth{0pt}

\begin{deluxetable}{ccccccccccccc}
\tabletypesize{\small}
\rotate
\tablecaption{Measured Abundances}
\tablehead{
\colhead{Star\tablenotemark{a}} 	& 
\colhead{[Fe/H]}      &
\colhead{[Fe II/H]}      &
\colhead{[Al/Fe]}      &
\colhead{[Ca/Fe]}      &
\colhead{[Sc II/Fe]}      &
\colhead{[Ti/Fe]}      &
\colhead{[Ti II/Fe]}      &
\colhead{[Cr/Fe]}      &
\colhead{[Co/Fe]}      &
\colhead{[Ni/Fe]}      &
\colhead{[La II/Fe]}      &
\colhead{[Eu II/Fe]}	
}
\startdata
III$-$53	&	$-$1.44	&	\nodata	&	0.34	&	0.24	&	$-$0.20	&	0.20	&	\nodata	&	\nodata	&	0.15	&	0.15	&	\nodata	&	\nodata	\\
III$-$85	&	$-$1.44	&	$-$1.40	&	0.06	&	0.37	&	$-$0.10	&	0.15	&	0.29	&	\nodata	&	0.07	&	0.04	&	0.04	&	0.44	\\
IV$-$30	&	$-$1.49	&	$-$1.45	&	0.26	&	0.49	&	0.02	&	0.29	&	0.20	&	0.01	&	\nodata	&	0.01	&	0.39	&	0.50	\\
IV$-$49	&	$-$1.49	&	$-$1.54	&	0.43	&	0.52	&	0.16	&	0.31	&	0.37	&	\nodata	&	\nodata	&	0.14	&	\nodata	&	0.67	\\
IV$-$86	&	$-$1.40	&	$-$1.43	&	0.56	&	0.50	&	0.29	&	0.18	&	0.24	&	0.19	&	0.27	&	0.09	&	$-$0.18	&	0.56	\\
\hline
\multicolumn{13}{c}{Cluster Mean Values}    \\
\hline
$\langle$$\rangle$	&	$-$1.45	&	$-$1.46	&	0.33	&	0.42	&	0.03	&	0.23	&	0.28	&	0.10	&	0.16	&	0.09	&	0.08	&	0.54	\\
$\sigma$	&	0.04	&	0.06	&	0.19	&	0.12	&	0.19	&	0.07	&	0.08	&	0.12	&	0.10	&	0.06	&	0.29	&	0.10	\\
$\pm$	&	0.02	&	0.03	&	0.08	&	0.05	&	0.09	&	0.03	&	0.04	&	0.09	&	0.06	&	0.03	&	0.16	&	0.05	\\

\enddata
\tablenotetext{a}{Identifies are from Arp (1955).}
\end{deluxetable}

\end{document}